\documentclass[10pt,a4paper]{article}
\usepackage[utf8]{inputenc}
\usepackage[english]{babel}
\usepackage{amsmath}
\usepackage{amsfonts}
\usepackage{amssymb}
\usepackage{graphicx}
\usepackage{fullpage}
\usepackage{comment}
\usepackage{hyperref}
\usepackage{indentfirst} 
\usepackage{authblk} 
\usepackage{comment}

\newcommand{\rieff}{R^{\text{eff}}_{i}}
\newcommand{\reff}{R^{\text{eff}}}

\newcommand{\rvec}{ \mathbf{r} }
\newcommand{\Rvec}{ \mathbf{R} }
\newcommand{\fvec}{ \mathbf{f} }

\newcommand{\vvec}{ \mathbf{v} }

\title{Phase Behavior  of Block Copolymer Nanocomposite Systems}

\author[1]{Javier Díaz}
\author[1]{Marco Pinna\thanks{mpinna@lincoln.ac.uk}}
\author[1]{Andrei V. Zvelindovsky}
\author[2]{Ignacio Pagonabarraga\thanks{ipagonabarraga@ub.edu}}
\affil[1]{School of Mathematics and Physics, University of Lincoln. Brayford Pool, Lincoln, LN6 7TS, UK}
\affil[2]{Departament de Física de la Matèria Condensada, Universitat de Barcelona, Martí i Franquès 1, 08028 Barcelona, Spain
}

\begin{document}
\maketitle

\begin{abstract}
    Nanocomposite materials made of block copolymer and nanoparticles display properties which can be different from the purely polymeric matrix. 
    The resulting material is a crossover of the original properties of the block copolymer and the presence of the assembled nanoparticles. 
    We propose a mesoscopic study via cell dynamic simulations to quantitatively assert the  properties of such hybrid materials. 
    The most relevant parameters are identified to be the  fraction of nanoparticles in the system and its chemical affinity, while the nanoparticle size with respect to the block copolymer length scales plays a role in the assembly. 
    The morphological phase diagram of the block copolymer is constructed for nanoparticles with chemical affinity ranging from A-compatible to B-compatible for a symmetric A-B diblock copolymer.
    Block-compatible nanoparticles is found to induce a phase transition due to changes in the effective concentration of the hosting phase, while interface-compatible particles induces the appearance of two new phases due to the saturation of the diblock copolymer interface.
\end{abstract}

\section{Introduction}

The introduction of nanoparticles (NPs) into polymeric matrices has proved to result in better performing materials, such as the well-known case of carbon-black\cite{chodak2001relation}. In the case of block-copolymer/ nanoparticle mixtures, adding  colloids can reduce the critical electric field needed to induce alignment of the domains with the external electric field\cite{liedel2013electric}. 
Block copolymer (BCP) materials are strongly periodic, serving as perfect templates for controlled placement of colloids\cite{okumura2000nanohybrids,kim2006effect}, resulting in highly ordered materials. 
In order to achieve a detailed control of the NP assembly within the BCP matrix it is necessary to be able to predict the phase behavior  of the overall system once nanoparticles are added, as the distribution of nanoparticles  critically depends on the morphology of the BCP, and the morphology itself is modified by the presence of NPs. We aim to provide a clear picture of the phase behavior  of diblock copolymer systems in the presence of nanoparticles.

Modifying the surface of the nanoparticles to make them compatible with one of the blocks can lead to a precise localization of the NPs within BCP domains \cite{chiu2005control,chiu2007distribution,horechyy2014nanoparticle} or the interface between them\cite{kim2006effect,kim2007creating}. Since the presence of nanoparticles can induce a phase transition of the BCP, it is crucial to determine the overall morphology of the polymer nanocomposite system. We explore the phase diagram of several parameters (both polymeric and colloidal) in order to establish a clear picture of the interplay between the pure BCP matrix and the colloidal fillers depending on the BCP architecture, the nanoparticle shape and interactions, both between colloids and with respect to the BCP. 

BCP/NP systems have been widely studied with theoretical and computational techniques. 
Strong segregation theory have been used to analitically study the viscoelastic propeties of polymer nanocomposites
\cite{kim2002morphology,pryamitsyn2006strong,pryamitsyn2006origins}
, finding a reduction of the lamella thickness when non-selective nanoparticles are present in the interface. A lamellar to bicontinuous transition was also reported, given by the vanishing of the bending modulus of the diblock copolymer, which is in accordance with experimental findings \cite{kim2007creating}.  

Simulation techniques such as Monte Carlo have been used \cite{detcheverry2008monte,kang2008hierarchical} to assert the assembly of BCP/NP systems on chemically nanopatterned substrates. In close resemblance with experiments, this method allowed to obtain well-ordered assembled nanoparticles. Furthermore, Huh et al\cite{huh2000thermodynamic} reported the changes in the diblock copolymer morphologies due to the presence of A-compatible nanoparticles in a diblock copolymer of arbitrary morphology (that is, exploring the composition ratio) using 3D simulations. This provided a phase diagram with only a few points. Molecular Dynamics \cite{schultz2005computer} was used to study the phase behavior of BCP/NP systems for different Flory–Huggins parameter values using fixed symmetric diblock copolymers. It was also reported that nanoparticle localisation is increased with nanoparticle size, as was experimentally found by Bockstaller et al \cite{bockstaller2003size}.

The aggregation of two kinds of nanoparticles within diblock copolymer was examined using Dissipative Particle Dynamics \cite{liu2006cooperative,chen2010structure,posocco2010molecular,maly2008self}, finding NP-assembly dependence on the lamella morphology, resulting in a transition to a complex phase. These simulation methods often require considerable computational time, thus limiting the simulation box to less than three or four periodic domains. 

Self Consistent Field Theory
\cite{thompson2001predicting,thompson2002block,matsen2008particle,lee2002effect,ginzburg2005influence}
has been widely used to study the segregation of nanoparticles within the diblock copolymer domains, again reporting the size-selectivity of NP localisation found in experiments\cite{bockstaller2003size}. 
Using SCFT Ginzburg et al \cite{ginzburg2005influence,he2006determining} explored the order-disorder transition of a binary polymer blend in the presence of A-compatible nanoparticles. While Lee et al\cite{lee2002effect} provided phase diagrams of the diblock copolymer morphology in terms of the composition ratio and the affinity of the nanoparticle, the volume fraction of nanoparticles was not explored. For this reason, we will consider a varying number of particles, on top of different NP chemical coating.  
 The Cahn-Hillard equation \cite{balazs2000multi,ginzburg2000modeling,ginzburg2002three} has been used to study the dynamical evolution of the phase separation, which is found to be slowed down by the presence of nanoparticles in the polymer blend. In these cases, a moderate volume fraction of nanoparticles that do not interact with each other is considered. 
 
While these works have addressed diblock copolymer nanocomposites, only Huh et al \cite{huh2000thermodynamic} has provided a quantitative study of the phase behavior of the diblock copolymer in the presence of a concentration $\phi_p$ of nanoparticles, with a limited amount of points. Furthermore, we aim to make use of computationally inexpensive methods to study such systems in detail and being able to tackle a large variety of parameters. For instance, we do not restrict our study to A-compatible colloids. By doing so, we are considering a more general approach than previous works, focusing on the morphologies of the resulting nanocomposite system, while a future work will tackle the assembly of colloids, under the same principles. Similarly, we do not restrict to a particular BCP matrix, i.e. we study a range of values of composition ratio $f_0$. This allows to predict with generality the resulting phase of a diblock copolymer nanocomposite.  

Here we expand our previous work in which NP-induced morphological transitions in the BCP were reported \cite{pinna2011modeling,diaz2017cell}. We aim to obtain a quantitative picture of the morphological behavior  of BCP/NP systems using a hybrid in-grid Cell Dynamic Simulation method along with an out-of-grid Brownian Dynamics for the BCP and for the NP, respectively. The coarse grained nature of the Cahn-Hillard equation suits this approach, as relatively large systems (in terms of several domain periods) are not numerically expensive. The model will be briefly presented in section \ref{sec:model} while results will be shown in section \ref{sec:results}. Throughout this work we are restricted to 2D simulations which allow to perform a large number of simulations with considerable system sizes. The basic features of this study are  captured in two dimensions, while a future work will cover three dimensional simulations which are particularly interesting in the case of bicontinuous phases.  

\section{Model}\label{sec:model}

Here we present a  coarse grained model in which the BCP is described as differences in concentration, hence, the individual monomers are not resolved. The reason for this choice is that we are more interested in the morphological properties of the diblock copolymer, rather than the microscopic properties of the polymer chain. Nanoparticles are individually treated and interact with the scalar field that stands for the BCP, resulting in a hybrid particle/field model\cite{tanaka2000simulation}. This treatment allows us to consider a large number of BCP periods, which is needed to study the morphology of BCP/NP systems without a huge computational effort. 

The BCP is characterized by the order parameter $\psi(\rvec,t)$ which is related to the differences in the local monomer concentration $\phi_A(\rvec,t)$ and $\phi_B(\rvec,t)$ of block A and B, respectively, 
\begin{equation}
\psi(\rvec,t)=\phi_A(\rvec,t)-\phi_B(\rvec,t)+(1-2f_0)
\end{equation}
with the composition ratio $f_0=N_A/(N_A+N_B)$ being the overall volume fraction of monomers A in the system. $\psi(\rvec,t)$ is considered the local order parameter, which has a value $0$ for the disordered-or homogeneous- state and $|\psi|>0$ for microphase-separated regions. 

The time evolution of $\psi(\rvec,t)$ is dictated by the conservation of mass, resulting in the Cahn-Hilliard-Cook equation \cite{cahn1958free,cook1970brownian}
\begin{equation}
\frac{\partial\psi ( \rvec, t )}{\partial t}=
M \nabla^2 \left[
\frac{\delta F_{tot} [ \psi] }{ \delta \psi}
\right]+
\eta ( \rvec, t)
\label{eq:cahn}
\end{equation}
with $M$ being a mobility parameter and $\eta(\rvec,t)$ being a gaussian noise parameter that satisfies the fluctuation-dissipation theorem
\begin{equation}
\langle \eta(\rvec,t) \eta(\rvec',t')\rangle =
-k_B T M \nabla^2 \delta(\rvec-\rvec')
\delta(t-t')
\end{equation}
for which we have used the algorithm given by Ball\cite{ball1990spinodal}. $k_BT$ sets the thermal energy scale of the diblock copolymer. 

The total free energy present in Equation \ref{eq:cahn} is decomposed into purely polymeric, coupling and intercolloidal free energy, respectively, 
\begin{equation}
F_{tot}=
F_{OK}+F_{cpl}+F_{cc}
\end{equation}
where the purely polymeric free energy $F_{OK}$ is the standard Ohta-Kawasaki free energy \cite{ohta1986equilibrium}. Furthermore, the diblock copolymer free energy $F_{OK}=F_{sr}+F_{lr}$ can be decomposed in short ranged
\begin{equation}
\label{eq:Fshort}
F_{\text{sr}}[\psi]=\int d\rvec 
\left[ 
H(\psi)+\frac{1}{2} D |\nabla\psi|^2 
\right]
\end{equation}
and long-ranged free energy, 
\begin{equation}
\label{eq:Flong}
F_{lr}[\psi]=
\frac{1}{2} B\int d\rvec \int d\rvec'
G(\rvec,\rvec')\psi(\rvec)\psi(\rvec')
\end{equation}
with $G(\rvec,\rvec')$ satisfying $\nabla^2 G(\rvec,\rvec')=-\delta(\rvec-\rvec')$,i.e., the Green function for the Laplacian. 

The local free energy can be written as  \cite{hamley2000cell}
\begin{equation}
H(\psi)=
\frac{1}{2}\tau'\psi^2 
+\frac{1}{3} v(1-2f_0)\psi^3 +\frac{1}{4} u \psi^4
\end{equation}
where $\tau'=-\tau+A(1-2f_0)^2$, $u$ and $v$ can be related to the molecular structure of the diblock copolymer chain \cite{ohta1986equilibrium}. 
The local free energy $H(\psi)$ possesses 2 minima values $\psi_{-}$ and $\psi_{+}$ which are the values that $\psi(\rvec,t) $ takes in the phase-separated domains.  
Parameter $D$ in Equation \ref{eq:Fshort} is related to the interface size $\xi=\sqrt{D/\tau'}$ between domains and $B$ in Equation \ref{eq:Flong} to the periodicity of the system $H\propto 1/\sqrt{B}$ as the long ranged free energy takes into account the junction of the two chains in a diblock copolymer.

Contrary to the block copolymer -which is described continuously- NPs are individually resolved. 
We consider a suspension of $N_p$ circular colloids with a tagged field moving along its center of mass $\psi_c(r)$.
 The presence of  nanoparticles in the BCP is introduced by a coupling term in the free energy,  which takes a simple functional form
\begin{equation}
F_{cpl}[\psi,\{\Rvec_i\}]= 
\sum_{p=0}^{N_p}
\sigma\int d\rvec\ \psi_{c}\left(\rvec-\Rvec_p \right)
\left[\psi(\rvec,t)-\psi_0    \right]^2
\end{equation}
with $\sigma$ a parameter that controls the strength of the interaction and $\psi_0$ an affinity parameter that is related to the preference of the NP towards different values of the order parameter $\psi(\rvec,t)$. A particle with an affinity $\psi_0=1$ is purely coated with copolymer A while a mixed brush would result in $\psi_0=0$. 
$\psi_c(\rvec)$ is a tagged function that accounts for the size and shape of the nanoparticle. 
At the same time, it can be tuned to define a soft and a hard-core for the nanoparticle regarding the coupling with the BCP. 
In our simulations we use 
\begin{equation}
\psi_{c} (\rvec) = 
\exp\left[
1-\frac{1}{1-\left( \frac{| \rvec   |}{\rieff}  \right)^\alpha} 
\right]
\label{eq:psici}
\end{equation}
from which we obtain a relationship $\reff = R_0  \left( 1+1/\ln 2    \right)^{1/\alpha} $ such that the tagged field has been reduced to $0.5$ at $r=R_0$. $R^{eff}$ also acts as the cut-off distance in the coupling interaction, i.e., $\psi_c(r>R^{eff})=0$

Nanoparticles are considered soft in their interparticle interaction, following a Yukawa-like potential 
\begin{equation}
U(r)=
U_0 
\left[\frac{ 
\exp\left(1-r/R_{12}   \right)}{r/R_{12}}-1
\right]
\end{equation}
with $R_{12}=2R_0$ and $r$ being the center-to-center distance. When an attractive component is desirable a standard Lennard-Jones potential has been used. 

 Colloids undergo diffusive dynamics, described by the Langevin equation in the over damped regime. The center of mass of each colloid $\Rvec_i$ is considered to follow Brownian Dynamics, that is,
\begin{equation}
\label{eq:brownian}
\vvec_i=
\frac{1}{\gamma} \left(
\fvec^{c-c}+\fvec^{cpl}+\sqrt{2k_BT\gamma}\xi
 \right)
\end{equation}
with $\gamma$  the friction coefficient, $k_BT$ is the NP thermal energy and $\xi$ is a random gaussian term satisfying fluctuation dissipation theorem. The coupling force $\fvec_i^{cpl}=-\nabla F_{cpl}$ accounts for the interaction between the nanoparticle and the BCP medium. 

The order parameter time evolution presented in Equation \ref{eq:cahn} is numerically solved using a cell dynamic simulation scheme\cite{oono1988study,bahiana1990cell}, for which the laplacian is approximated as $\frac{1}{a_0^2}  [ \langle\langle X \rangle\rangle -X  ] $ with 
\begin{equation}
\langle\langle \psi \rangle\rangle = \frac{1}{6}  \sum_{NN}  \psi   +\frac{1}{12} \sum _{NNN} \psi
\end{equation}
in two-dimensional systems. NN and NNN stand for nearest-neighbour and next-nearest-neighbour, respectively, that is, summation over lattice points around the lattice point $\psi_{ij}$. 
The lattice is characterized by its spacing $a_0$.  

In all cases $\psi(\rvec,t=0)$ is randomly distributed, corresponding to a disordered phase. In addition, the initial state for the colloidal center of mass is randomly chosen, with the condition of non-overlapping each other. The system is then let to evolve following the dynamical Equations \ref{eq:cahn} and \ref{eq:brownian} until a stationary state is approximately reached. 
Although a true equilibrium profile can not be assured, the time evolution of the  microphase separation of the diblock copolymer can be tracked by $\langle | \psi(\rvec,t) | \rangle$, as shown by \cite{ren2001cell}.    
A typical simulation run of a system  sized $V=256\times 256$ requires a few hours of serial computational time. 

In summary, we make use of a hybrid model that combines a continuous description of the diblock copolymer that has been widely used to describe its phase separation process, along with an individual characterization of each colloidal particle.

\section{Results}\label{sec:results}

The aim of this work is to study the phase behavior  of a diblock copolymer mixture in the presence of  nanoparticles. For that purpose we restrict our study to a series of parameters which will be presented in the form of phase diagrams. The overall concentration of particles is the fundamental variable that will be considered, hence the purely polymeric properties can be retrieved at $\phi_p=0$.  Exploring a vast range of $\phi_p$ allows to consider several regimes, as we will consider both low-concentration and high concentration regimes of nanoparticles. 

Since our objective is to assert composite systems, the purely polymeric parameters such as $\chi N$ are not explicitly explored. Instead, we will tune $f_0$ to study different morphologies and therefore we are approximately covering horizontal lines in the well-known $\chi N - f_0$ phase diagram\cite{matsen1996unifying}.  Similarly, domain size is explored in comparison with the nanoparticle size. Then, we use fixed values for the BCP parameters, which are standard in the literature\cite{ren2001cell,pinna2012large}: $A=1.5,D=0.5,\tau=0.35,v=1.5,u=0.5$. When used, the noise in the Cahn-Hillard equation has a strength of $\xi=0.1$. 

\begin{figure}[h!]
\centering
\includegraphics[width=0.7\textwidth]{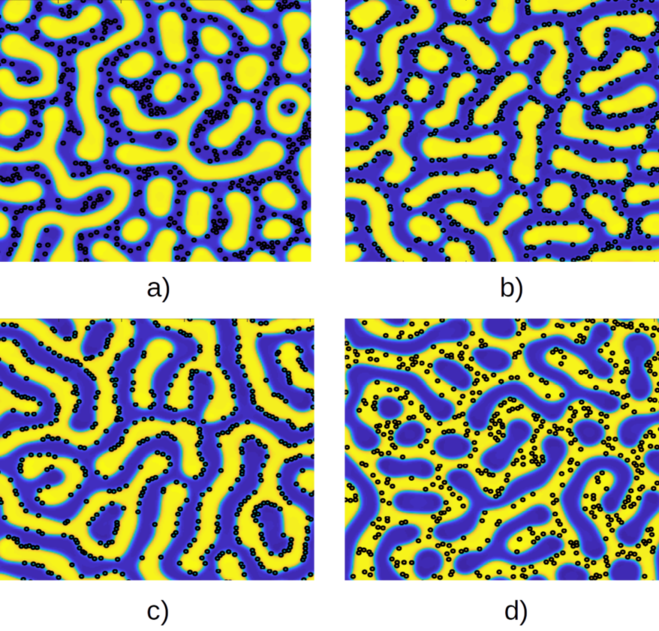}
\caption{Low volume fraction ($\phi_p\approx0.056$) segregation of NPs for $\psi_0=-1$ (a), $\psi_0=-0.33$ (b), $\psi_0=0$ (c) and $\psi_0=0.67$ (d), in a symmetrical lamellar matrix.
Yellow and blue regions stand for A-rich and B-rich monomer, respectively. Nanoparticles are shown as black circles. }
\label{fig:chem-snaps}
\end{figure}

We firstly propose to study the effect that a varying affinity $\psi_0$ of nanoparticles have on a given block copolymer morphology. In Figure \ref{fig:chem-snaps} we can see four final snapshots of a relatively low  concentration of nanoparticles ($5.6 \%$ of surface) dispersed in a symmetric lamellar-forming diblock copolymer. NPs are sized $R/L\approx 0.09$ with respect to the lamellar domain size (half of the period). In a) the affinity is $\psi_0=-1$ meaning that the particles are totally compatible with the blue phase, therefore all colloids  are well dispersed within its preferred phase. If the NPs are almost neutral they segregate to the interface (b), but they do only symmetrically when its affinity is $\psi_0=0$ (c). At $\psi_0=0.67$ (d), they are again dispersed within the yellow phase, but fewer NPs are found in the center of the domain, as the affinity is not identically $1$. 

Figure \ref{fig:chem-snaps} clearly displays the importance of the affinity $\psi_0$ in the localization of NPs. As experiments  have shown\cite{kim2007creating}, the surface of NPs can be coated to be compatible with a part of the A-B diblock copolymer. In practice, typical ligands will be an A-copolymer or B-copolymer brush, or a mixed one, resulting into A-selective, B-selective or neutral nanoparticles, respectively. In theoretical  works\cite{thompson2002block} this enthalpic interaction is introduced via two parameters $\chi_{AP}$  and $\chi_{BP}$. In our model $\psi_0$ plays the role of the difference in enthalpic interaction and $\sigma$ the NP/BCP interaction strength. 

\begin{figure}[h!]
\centering
\includegraphics[width=0.9\textwidth]{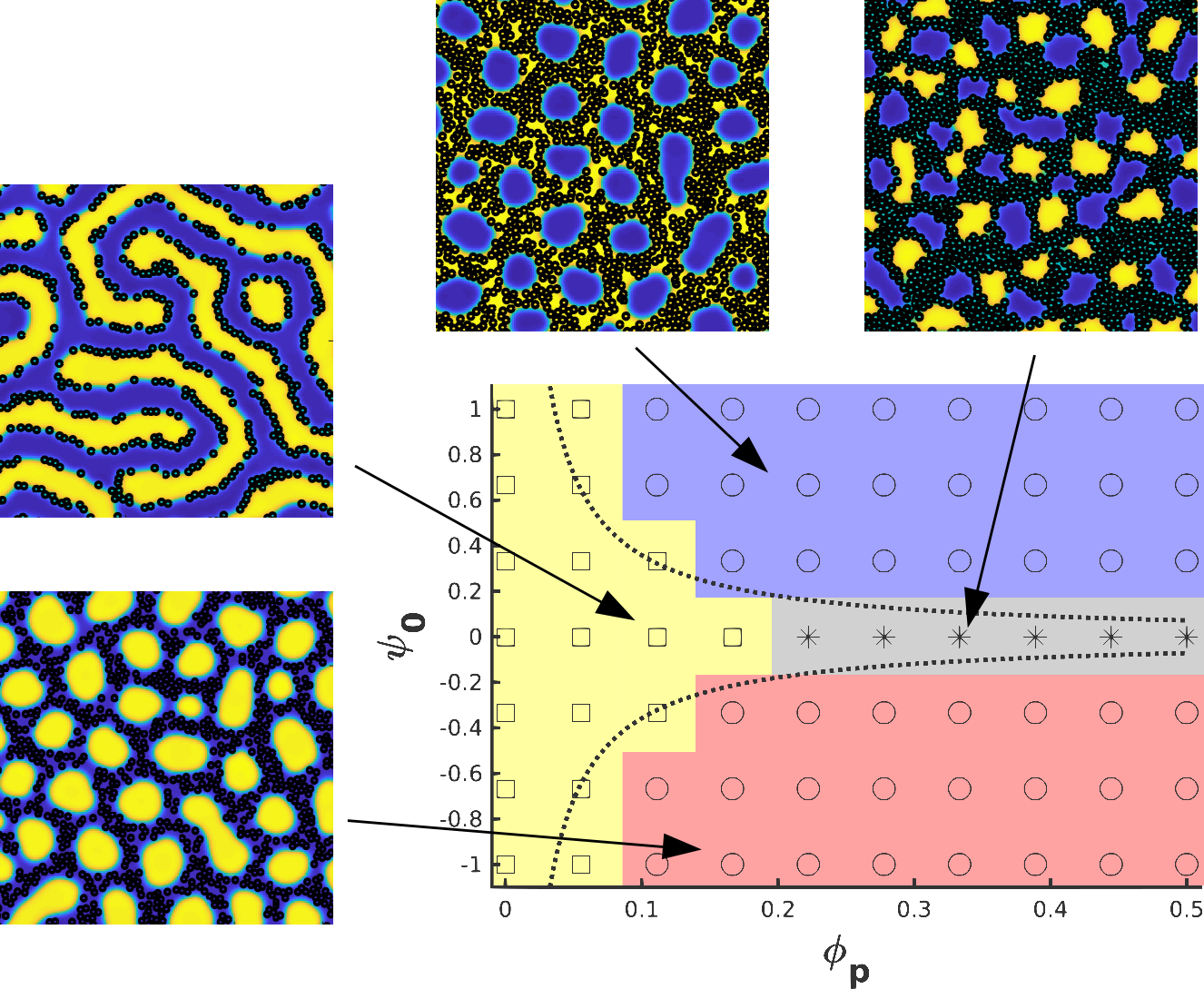}
\caption{Phase diagram of a symmetric BCP with a a volume fraction $\phi_p$ of NPs with different affinity, given by $\psi_0$. Squares represent lamellar, circles stand for cylindrical phase while asterisk for broken-lamellar phase. The dashed curve represents Equation \ref{eq:feff}. }
\label{fig:chem}
\end{figure}

Furthermore, we can explore a range of concentrations of colloids so that the presence of nanoparticles can be quantified. In Figure \ref{fig:chem} we have simulated a range of NP concentrations $\phi_p$ and several values of colloidal affinity $\psi_0$. At low concentration $\phi_p<0.1$ nanoparticles are simply placed within the preferred region, therefore the morphology of the BCP remains  lamellar, represented by squares in the yellow region. 
 
 We observe that selective particles  ($|\psi_0|=1$) induce a phase transition from lamellar to cylindrical due to the effective increase in the composition ratio of the hosting domain. These two regions are symmetrically placed around the $\psi_0=0$ line, as blue and red regions.Contrary to that, in the $\psi_0\approx 0$ regime NPs segregate to the interface until a saturation volume fraction $\phi^*_p$ is reached (similarly to what we described in Figures \ref{fig:chem-snaps} b) and c) ). At higher number of particles the  domains are broken to create more interface. Eventually the lamella structure is destroyed to find alternating irregular domains surrounded by a continuous region of nanoparticles.

Intermediate values of the affinity result in particles that are neither totally compatible with one of the domains nor symmetrically placed in the interface. 
Therefore a larger number of particles is needed to induce the lamella-to-cylindrical phase transition. 
This accounts for the shape of the blue and red regions and can be quantitatively described by the expression,
\begin{equation}
f_{eff}=
\psi_0\phi_p+(1-\psi_0\phi_p)f_0
\label{eq:feff}
\end{equation}
which derives from the one introduced by Huh \cite{huh2000thermodynamic} after the simple substitution $\phi_p\rightarrow\phi_p \psi_0$. In Figure \ref{fig:chem} this equation is showed by the dashed line which approximately describes the behavior  of the phase transition. 

Figure \ref{fig:chem} explores several values of the affinity $\psi_0$ in a lamellar-forming BCP and two regimes have arisen: Compatible particles induce an order-to-order phase transition based on an increase of the effective concentration of the hosting phase. 
On the other hand, neutral nanoparticles break the BCP domain up by saturation of the interface. 
It can be noted that 3 different morphologies can be induced from a single lamellar-forming BCP matrix by a combination of NP loading and chemical coating of the NP's surface. 
We can now focus on these two regimes by fixing $\psi_0$ but explore the different morphologies of the BCP by changing $f_0$.

\begin{figure}[h!]
\centering
\includegraphics[width=0.9\textwidth]{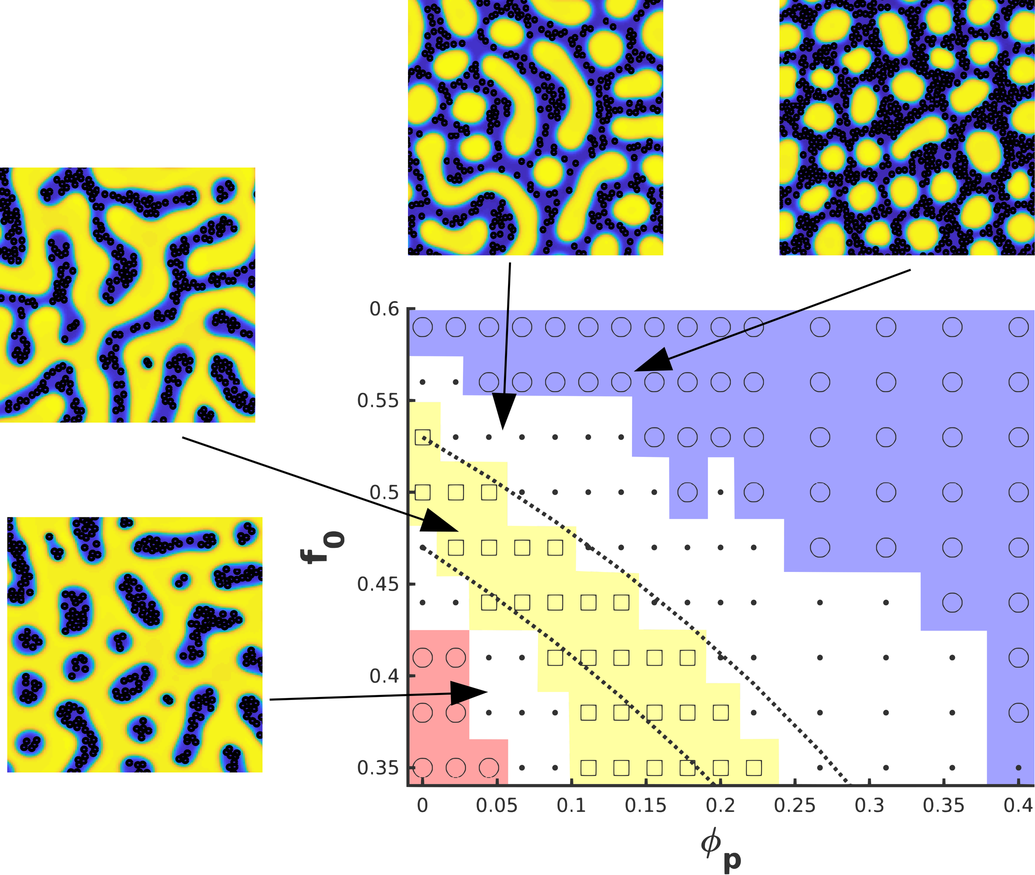}
\caption{Morphology of a BCP/NP system with a $f_0$ ratio and $\phi_p$ nanoparticles which are strongly compatible towards one of the phases. Shown phases are: Circular phase(Circles),Mixed phase (dots) and lamellar (squares).}
\label{fig:compatible}
\end{figure}

Firstly, we consider a volume fraction of NPs that are strongly compatible with one of the blocks by choosing $\psi_0=-1$. Various values of $f_0$ are considered so that we can observe the effect that NPs have on different BCP morphologies. NPs which are compatible with the majority phase will simply segregate within its hosting phase, which already percolates the system, therefore we are more interested in the case in which the hosting domain is the minority phase. For that reason in Figure \ref{fig:compatible} we explore $0.35<f_0<0.6$, that is, in the absence of nanoparticles we observe the well-known phases: cylindrical ( red and blue circles) and  lamellar(yellow squares). A transition region is labeled as mixed phase (dots)   . 

At low $\phi_p$ we simply observe segregation of NPs towards the preferred domain. As the concentration is increased, the nanoparticles induce an increase in the effective volume fraction of the hosting phase, which can be tracked by Equation \ref{eq:feff}. Indeed, in Figure \ref{fig:compatible} we observe the NP-induced phase transition, which is found to qualitatively agree with Equation \ref{eq:feff}. Selective-NP-induced phase transitions have been observed both experimentally\cite{halevi2014co,kim2005nanoparticle,lo2007effect,
lo2010effect,kim2016morphology} and theoretically\cite{sides2006hybrid}. We have obtained a phase diagram with considerably more number of points than previous works\cite{huh2000thermodynamic}, which allows us to confine the regions for each morphology more precisely. For completion, it should be straight-forward to revert Figure \ref{fig:compatible}  for affinities $\psi_0=+1$, that is, nanoparticles with an affinity towards the other domain: the phase space would be symmetrical with respect to the $f_0=1/2$ line. 

\begin{figure}[h!]
\centering
\includegraphics[width=0.7\textwidth]{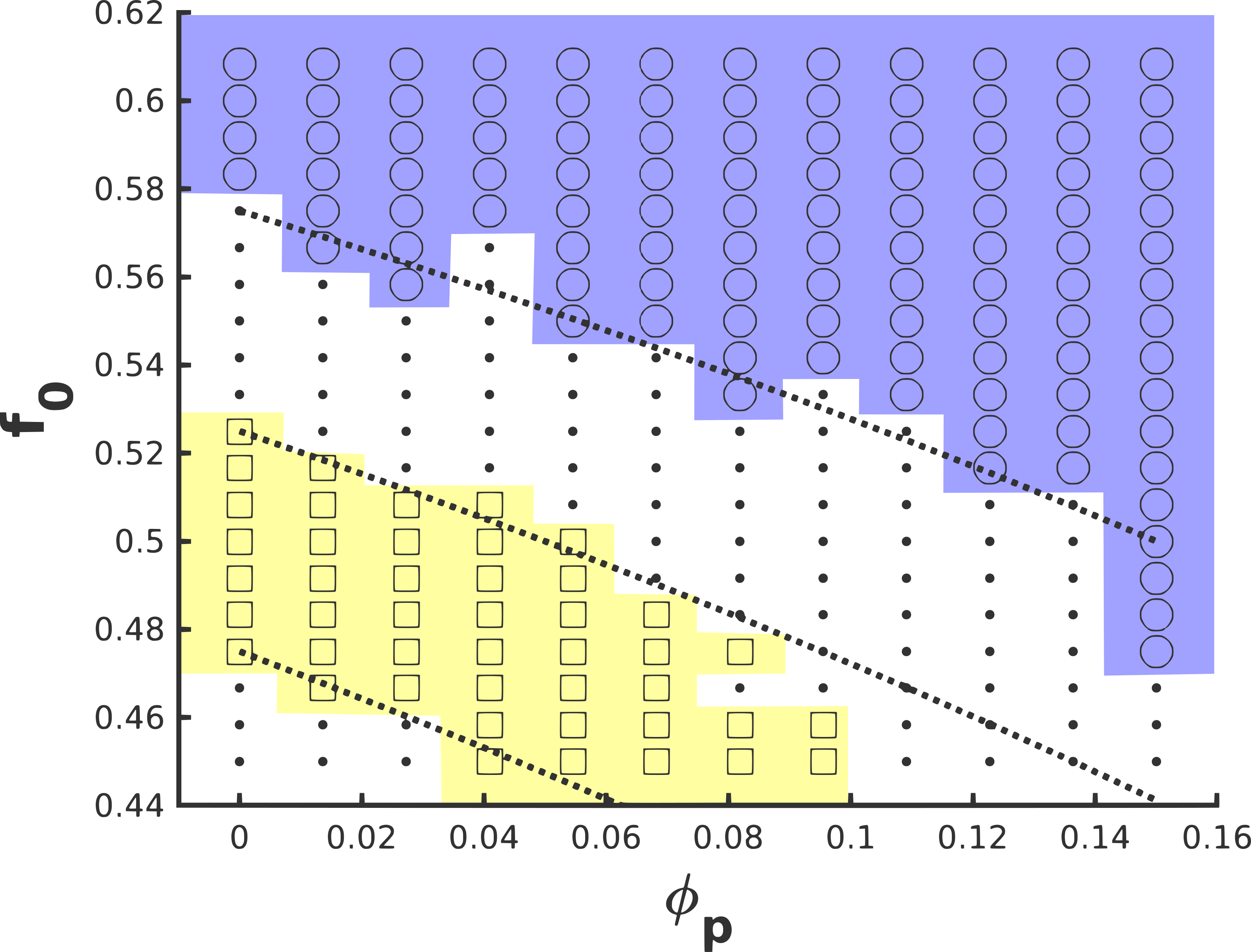}
\caption{Morphology of a BCP/NP system with a $f_0$ ratio and $\phi_p$ nanoparticles which are strongly compatible towards one of the phases. Shown phases are: Circular phase(Circles),Mixed phase (dots) and lamellar (squares).}
\label{fig:compatible.inset}
\end{figure}

While Figure \ref{fig:compatible} explores a vast range of number of particles in the system, we can focus on $\phi_p<0.16$ to study in detail the boundaries of the phase regions. To quantitatively determine the phase of the system with precision we ran $5$ simulations with the same parameters but different initial condition. Then the number of domains in the system was averaged to determine the phase of the system with considerable precision. The results are shown in Figure \ref{fig:compatible.inset}, where again NP-induced transitions are observed. The quantitative agreement between the simulations and Equation \ref{eq:feff} is considerable for small $\phi_p$. In fact, the disagreement between the phase diagram and the simple $f_{eff}$ expression can be easily tracked as follows: at low volume fraction nanoparticles are relatively isolated within their preferred domain, that is, they rarely interact with each other. As the nanoparticles start to fill the domains collective behavior  between nanoparticles appears, inducing an effective volume fraction that is larger than the ideal, described by Equation \ref{eq:feff}. Since the available volume for nanoparticles depend on the morphology of the system we can define an effective volume fraction $\phi_p^{eff}\equiv \phi_p/f_{eff}$ that explains why disagreeing regions appear at higher NP concentration as $f_0$ grows.  

From these simulations and Equation \ref{eq:feff} it is clear that $\phi_p$ is the most important magnitude in play when NPs are compatible with one phase. 
We aim to get a better insight of the phase behavior  of BCP/NP systems by using a different interparticle potential.
 We repeated the procedure described for Figure \ref{fig:compatible} using colloids that interact through a Lennard-Jones potential  and tuned the cut-off radius to set attractive and repulsive potentials. 
 A large number of simulations were performed and its result can be seen in Figure S1 in the Supporting Information with no difference found between attractive or repulsive particles. 
 Therefore, we can conclude that the interparticle potential does not play a role in the NP-induced phase transition with our level of precision.
 Nonetheless, the assembly of nanoparticles within their preferred phase is totally different depending on the kind of potential we use, having no consequence over the BCP morphology.

\begin{figure}[h]
\centering
\includegraphics[width=0.6\textwidth]{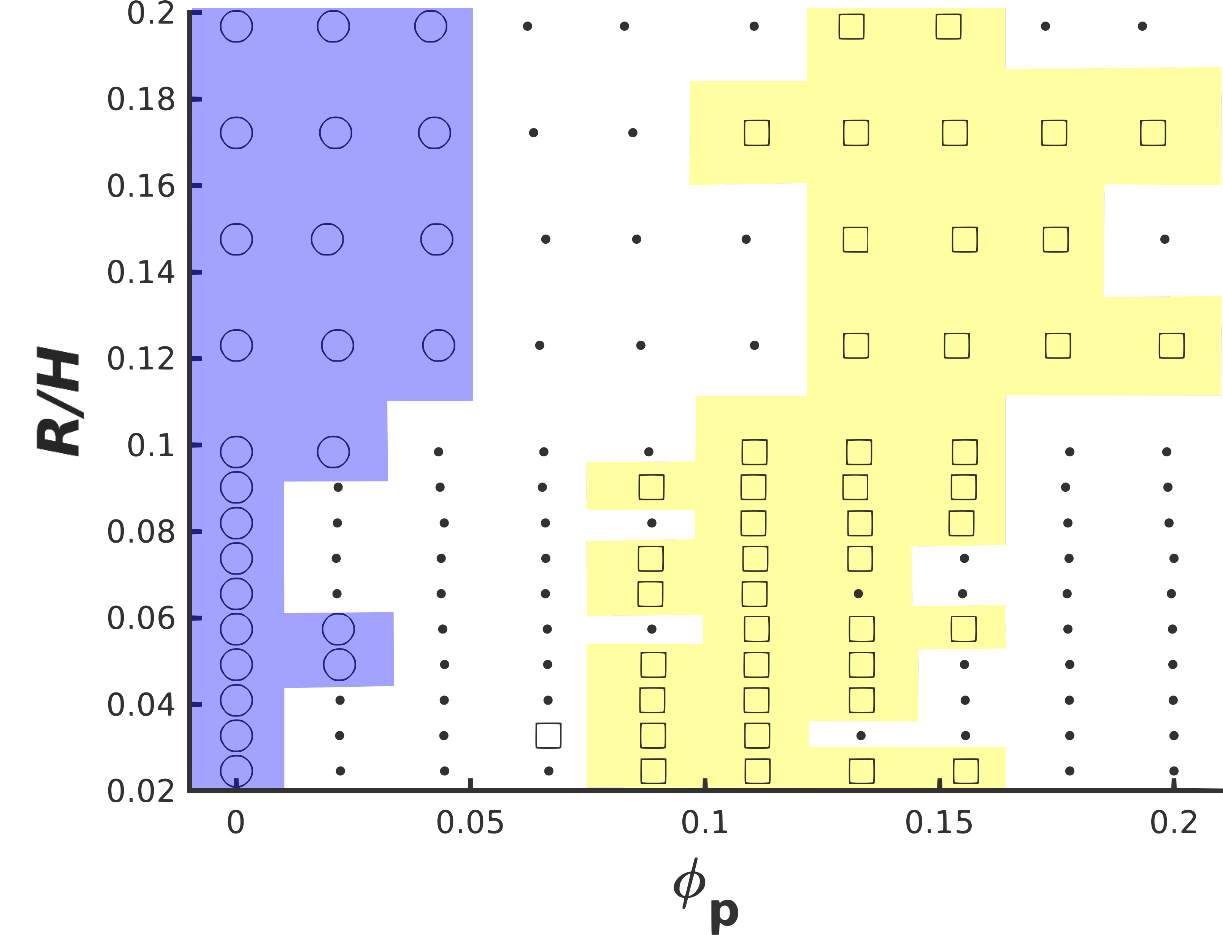}
\caption{Phase diagram of an asymmetrical BCP matrix ($f_0=0.4$) with different NP fraction $\phi_p$ of different sizes $R$, reduced with the size of the periodicity $H$ of the BCP morphology. The phases shown are: Cylindrical (Circles), Mixed (dots) and lamellar (Squares)}
\label{fig:sizes.comp}
\end{figure}

Similarly to the interparticle potential, the size of the nanoparticles is determinant in the assembly of nanoparticles but at first sight it is not expected to be influencial in the BCP morphology. 
To assert it we performed simulations of different volume fractions of compatible particles $\psi_0=-1$ for differently sized NPs. 
Small and large nanoparticles disturb the surrounding polymer in different ways, in fact, Figure \ref{fig:sizes.comp} shows that smaller nanoparticles with respect to the domain size are more effective at inducing phase transitions. 
This behavior  can be explained within our model as the characteristic length of the interface $\xi=2\sqrt{D/\tau}$ relates to the size of the nanoparticle: a large number of closely separated nanoparticles with a typical interparticle distance $d<\xi$ does not allow for the order parameter to restore small disturbances in the otherwise flat profile. Since no attractive interaction is present, this result in groups of NPs that are only loosely together. In terms of polymeric chains we can conclude that NPs can create NP-rich regions that result in an effective higher concentration of NPs.  


In Figure \ref{fig:chem} we found that neutral nanoparticles segregate to the interface and induce a phase transition which is essentially different to the one described in Figure \ref{fig:compatible}. Since that analysis was performed for a given BCP morphology we now aim to study the case of changing $f_0$.  Again, we now fix the affinity of nanoparticles and study the phase of  a block copolymer with a morphology given by $f_0$, in the presence of a number of particle $\phi_p$- using neutral nanoparticles instead. Therefore $\psi_0=0$, that is, nanoparticles always have a preference towards being symmetrically placed at the interface. 

\begin{figure}[h]
\centering
\includegraphics[width=0.8\textwidth]{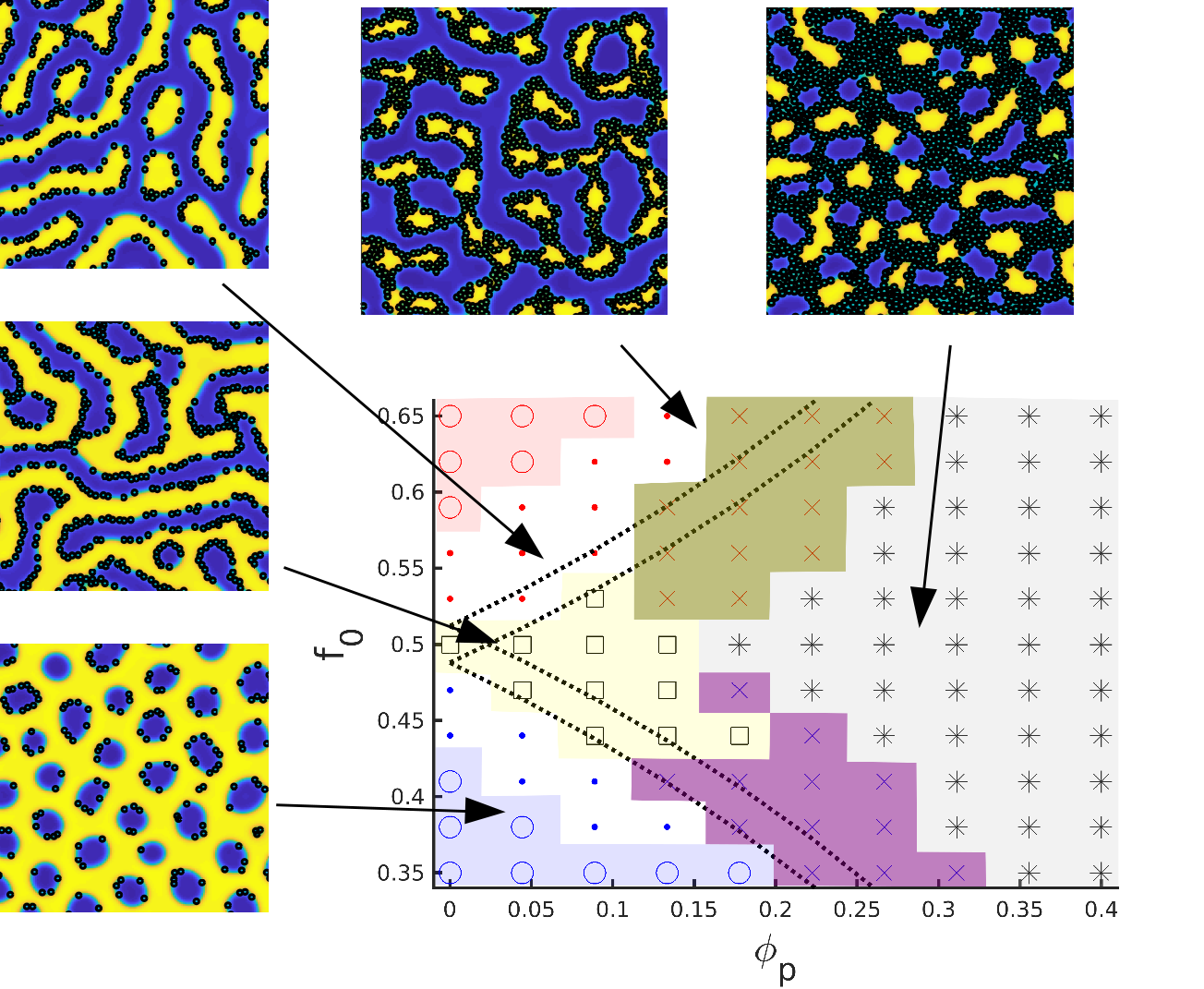}
\caption{Mophological phase diagram of neutral nanoparticles ($\psi_0=0$) in several diblock copolymer matrices given by $f_0$.  Shown phases are: Circular phase(Circles),Mixed phase (dots), lamellar (squares), broken lamella ($*$) and balanced lamella+circles ($\times$)}
\label{fig:neutral}
\end{figure}

\begin{figure}[hbtp]
\centering
\includegraphics[width=\textwidth]{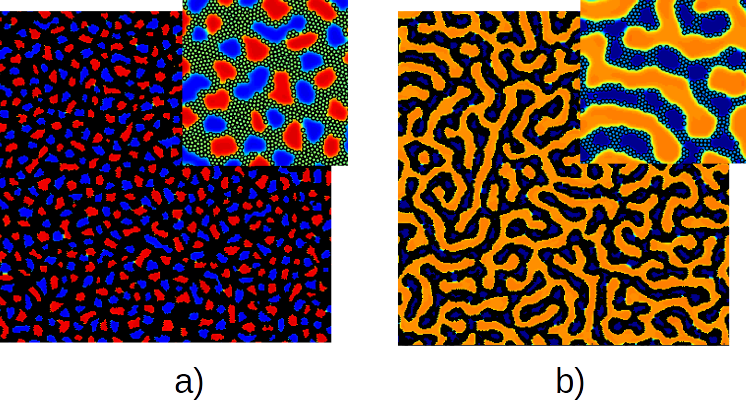}
\caption{Final snapshots of $512\times 512$ simulations with insets showing a $128\times 128$ subregion. In a) $\phi_p=0.4$ and $f_0=0.5$ while in b) $\phi_p=0.22$ and $f_0=0.35$   } 
\label{fig:newphases}
\end{figure}

At low volume fraction, in figure \ref{fig:neutral} we simply observe segregation of nanoparticles to the interface.
If we keep increasing the number of particles, a competition between NP-BCP coupling free energy and the $(\nabla \psi)^2$ term in Equation \ref{eq:Fshort} comes into action when the interface is filled with colloids. 
Therefore, new interface can be created when sufficient particles are present in the system. When the BCP's architecture is asymmetrical the presence of nanoparticles induces a transition from hexagonally-ordered  circular domains to elongated, which explains the presence of mixed-to-lamellar (dots to squares) and circular-to-mixed (circles to dots) phase transitions in Figure \ref{fig:neutral}. 
Symmetrical BCP (ie, lamellar, represented by squares) displays more interface in the absence of nanoparticles and therefore can accommodate a larger number of neutral nanoparticles than any other morphology. 
At sufficiently large number of particles lamellar domains are broken into smaller domains (asterisk). Figure \ref{fig:newphases} a) shows a large simulation of such a regime, in which the disordered overall structure can be seen.

At intermediate $\phi_p$ values we observe an interesting new phase (crosses) in the case of asymmetrical BCP (($\times$) in Figure \ref{fig:neutral}): in the presence of  a minority/majority BCP phase the NPs surround the minority domains and a particular balance $\phi_{majority}=\phi_{minority}+\phi_P$ can be reached, leading to the formation of a quasi-lamella phase for the majority phase and NP-enclosed minority domains, as can be also seen in Figure \ref{fig:newphases} b), where a considerably large system ($512\times 512$ grid points) are simulated. The overall orange phase percolates the surface, while blue domains are enclosed, as shown in the detail. 

This behavior  is not present any more when higher concentrations are reached (($*$) in Figure \ref{fig:neutral}). At this point the particular architecture of the BCP becomes less important, with alternating A-B domains dispersed into a NP-rich matrix. 

\begin{figure}[h]
\centering
\includegraphics[width=0.6\textwidth]{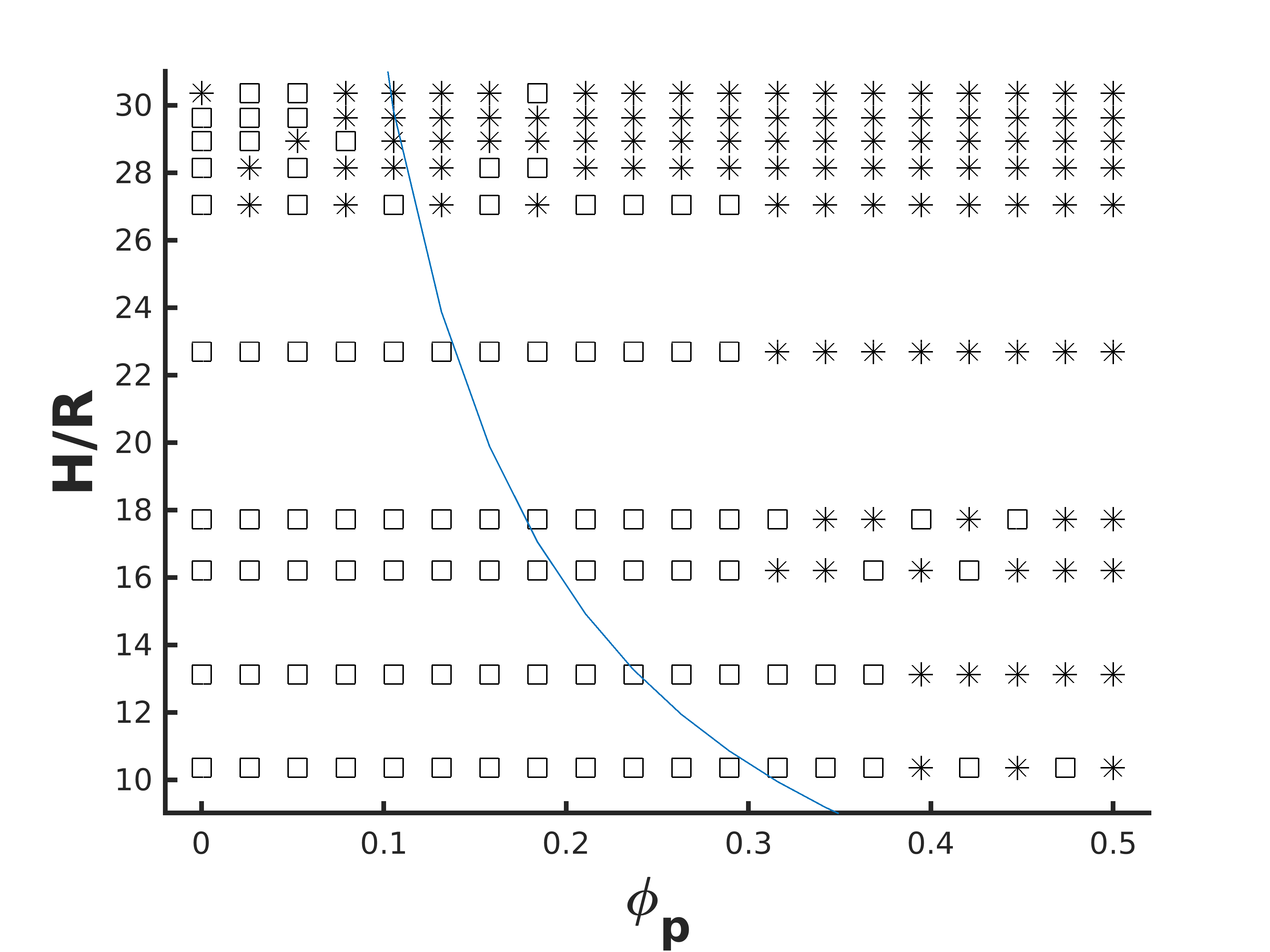}
\caption{Phase diagram of neutral nanoparticles in a lamellar BCP matrix of different domain periodicity ($H$) and at different NP radius ($R$). Square represent lamellar while asterisk stand for broken lamella. The blue curve shows a constant saturation line in accordance to Equation \ref{eq:gammap}   }
\label{fig:new_phase}
\end{figure}

The surfactant-like nature of neutral nanoparticles results in the volume fraction $\phi_p$ not being the most relevant parameter to account for the phase behavior . The fraction of the block copolymer interface that is occupied by nanoparticles is more relevant instead. This magnitude is not easily accessible as it is strongly dependent on the dynamical evolution of the system, therefore we consider the ratio between nanoparticles' linear dimension and interface size in the absence of colloids 
$
\gamma_p=
{2R N_p}/{\Gamma_0(f_0)}
$
. Ideally, in the symmetric BCP case the interface length only depends on the number of lamella periods that the system size can accommodate. Scaling with the size of the system $\Gamma_0(1/2)/V\sim 2/H$, with $H$ being the periodicity of the lamella. Therefore we can express the occupied interface fraction as 
\begin{equation}
\gamma_p\sim
\frac{\phi_p}{\pi R/H}
\label{eq:gammap}
\end{equation}
from which we can expect that the transition from lamella to broken-lamella phase will occur at $\phi_p^*\sim \pi\gamma_p^* \frac{R}{H}$ with $\gamma_p^*>1$ meaning that we are over the saturation point. 
Hence, smaller-sized lamellar can accomodate a larger number of particles and therefore a larger number of particles are needed to induce the breaking of the domains into smaller ones.  
We can confirm this by running simulations in which the size of the lamella domains is changed at different volume fractions and record the overall phase. In figure \ref{fig:new_phase} we can observe the different phases and conclude that Equation \ref{eq:gammap} approximately captures the trend of the two phases, although it is far from a quantitative agreement. This is firstly due to the interface length being a complicated parameter to predict, even when we only consider a pure polymeric system: the interface size is always larger than $2/H$, as the system will often be trapped in metastable states. Furthermore, colloids do not simply assemble side-by-side in the interface, instead colloids often form small local clusters. 

\section{Conclusions}

The phase behavior of BCP/NP composite systems has been studied in terms of the BCP composition and the nanoparticle loading, chemical affinity and size. The NP segregation within the BCP is  determined by the affinity of the particle, that is, the coating of the surface of each NP in a typical experimental setup. At low concentrations nanoparticles can either segregate to the interface between domains or be dispersed within its preferred phase. We have presented a phase diagram of the morphology of the diblock copolymer in terms of different values of the affinity. To our knowledge, this is the first time that the chemical affinity of the colloids has been studied, exploring totally different regimes ranging from A-compatible to B-compatible coating. 

In order to study high concentration cases with more generality we focus firstly on totally compatible particles, which segregate to one of the domains. 
Considering several BCP compositions we have found the expected order-to-order phase transition, and a phase diagram is constructed. This phase transition is due to an increase in the effective monomer fraction $f_{eff}$ of the hosting domain. 
The ability to tune the effective composition of a diblock copolymer with a given $f_0$ by adding compatible nanoparticles is an advantage, as one single BCP sample can lead to a variety of morphologies\cite{kim2016morphology}. 

The relatively fast CDS scheme has allowed to perform a large number of simulations to obtain a more precise phase diagram, in contrast with previous works using Monte Carlo methods \cite{huh2000thermodynamic}. 
Furthermore, we can compare the hybrid CDS/BD method with MC by plotting the transition curves given by Equation \ref{eq:feff} as described by Huh \cite{huh2000thermodynamic}. This confirms a good agreement with previous works. This $f_0-\phi_p$ phase diagram can be compared with experimental results via $f_{eff}$, as Lo et al.\cite{lo2007effect} directly mapped the morphology of a BCP/NP system with an effective composition ratio, obtaining a good agreement. 

The size of the nanoparticles with respect to the block copolymer lengths has been examined to find that the $R/\xi$ ratio plays a role in the morphological changes: smaller nanoparticles with respect to the interface size induces phase-transition more effectively. The interparticle potential has been shown not to affect the block copolymer, that is, the assembly of the nanoparticles does not influence the BCP morphology.

Secondly, neutral nanoparticles induce phase transitions in the BCP at concentrations higher than the saturation. In this regime NPs induce an increase in the BCP interface length, resulting in transitions towards elongated domains. The saturation process is strongly dependent on the size of the interface without colloids $\Gamma_0(f_0)$ which is larger for lamellar morphologies and smaller periodicity. The characteristic parameter is no longer the volume fraction of nanoparticles $\propto N_pR^2$ but an estimation of the fraction of nanoparticles in the interface $\propto N_p R$.  In the case of lamellar morphology, the saturation of the interface leads to domains being broken into shorter ones, again to induce higher interface length. At sufficiently large saturation rate the nanoparticles percolate the system, with alternating block copolymer domains dispersed within a colloid-rich matrix. We have been able to construct a phase diagram which states the various regimes that emerge when adding neutral nanoparticles to a diblock copolymer characterised by a composition ratio $f_0$. This new phase diagram allows to predict the morphologies of a diblock copolymer in the presence of neutral nanoparticles.  

Lamellar to bicontinuous morphological transitions have been reported in experiments using surfactant particles \cite{kim2009tailoring}. This behavior agrees qualitatively with our results involving a high concentration (above the saturation of the interface). Despite the simplicity of our model we are able to capture the same physical behavior as experiments. A future work will involve the precise comparison with experiments and theoretical prediction\cite{pryamitsyn2006strong}. 

Asymmetrical diblock copolymers in the presence of moderate fractions of particles adopt a NP-induced phase based on the balance of the majority phase with respect to the combination of minority phase and NP-rich regions. Given this balance, the majority block displays a lamella-like morphology while minority domains are enclosed by the nanoparticles. 

\nocite{langner2012mesoscale}
\nocite{bockstaller2005block}
\bibliographystyle{my_style4}
\bibliography{references}

\end{document}